\documentclass[twocolumn,prd,groupedaddress,nofootinbib,showpacs]
{revtex4}

\usepackage{latexsym}
\usepackage{amssymb}
\usepackage{graphicx}
\usepackage{dcolumn}
\usepackage{bm}

\begin{document}

\title{A Discussion on Supersymmetric  Cosmic Strings with Gauge-Field Mixing}
\author{C. N. Ferreira$^{1, 2, 3, a}$\thanks{crisnfer@if.ufrj.br}, C. F.
Godinho$^{2, 3, b}$\thanks{godinho@cbpf.br} and J. A. Helayel-Neto$^{2, 3, c}$\thanks{helayel@cbpf.br}}
\affiliation{$^1$Departamento de F\'{\i}sica, Universidade Federal do Rio de
Janeiro, POBOX 68528, 21945-910, Rio de Janeiro, RJ, Brazil
\\
$^2$
Centro Brasileiro de Pesquisas F\'{\i}sicas, Rua Dr. Xavier Sigaud
150, 
Urca 22290-180, Rio de Janeiro, RJ, Brazil
\\
$^3$Grupo de F\'{\i}sica Te\'orica Jos\'e Leite Lopes (GFT),
Petr\'opolis, RJ, Brazil
}

\date{\today}

\begin{abstract}
In this paper, following a stream of investigation on
supersymmetric gauge theories with cosmic string solutions, we
contemplate the possibility of building up a D-and-F term
cosmic string by means of a gauge-field mixing in connection with
a $U(1) \times U(1)'$-symmetry. The spontaneous break of both
gauge symmetry and supersymmetry are thoroughly analysed and the
fermion zero-modes are worked out. The role of the gauge-field
mixing parameter is elucidated in connection with the string
configuration that comes out. As an
application of the model presented here, we propose the
possibility that the supersimetric cosmic string yield
production of  fermionic charge carriers that may eject, at their
late stages, particles that subsequently decay to  produce cosmic
rays of ultra-high energy. In our work, it turns out that massive
supersymmetric fermionic partners may be produced for a susy
breaking scale in the range $10^{11}$ to $10^{13}$ GeV, which is
compatible with the phenomenology of a gravitino mass at the TeV
scale. We also determine the range of the gauge-field mixing
parameter, $\alpha $, in connection with the mass scales of the
present model.
\end{abstract}

\maketitle

\section{Introduction }

\noindent

Supersymmetry has been of interest to explain a number of problems
in Cosmology, inclu\-ding observations and astrophysical
questions. In this context,  supersymmetry may appear to offer a
version to acommodate the ``dark matter" problem. There are
evidences that   most  of the mass in the Universe is nonluminous
and of unknown composition, probably non-baryonic. The
supersymmetry framework predicts the existence of  new stable
elementary particles (neutralinos) having a mass less than a few
TeV and weak interactions with ordinary matter. The  neutralinos
are linear combinations of the SUSY partners of the photon
(photino), $Z^0$ and Higgs bosons. If such a weakly interacting
massive particle (WIMP) \cite{Jungman} exists, then it has a
cosmological abundance such as observed today, and we could
therefore account for the dark matter in the Universe.

Another important result is the success of duality in
supersymmetric Yang-Mills theories that may, by means of the physics of
non-perturbative solutions, such as topological solitons, be more natural
to understand than for non-supersymmetric theories. For these
reasons, supersymmetric extensions of cosmic string models are
especially important to our understanding of the early Universe.

In  Cosmology, the common belief is that, at high temperatures, symmetries that are
spontaneously broken today have already been
exact in the primordial stages of the Universe.
During the evolution of the Universe, there
were various phase transitions, associated with
the chain of spontaneous breakdowns of gauge symmetries.
Cosmic strings \cite{Vilenkin,Kibble1,Vilenkin1,Kibble2} are a
by-product of a series of symmetry breaking phase transitions
\cite{Kibble}, that appear in some GUT's and carry a
huge energy density. They may also
 carry enormous currents\cite{Witten} and  may provide a
sensible justification for many  astrophysical phenomena, such as the origin
of primordial magnetic fields \cite{Vachaspati}, charged vacuum
condensates \cite{Nascimento} and sources of ultra-high energy
cosmic rays \cite{Hill,Bhattacharjee1,Bhattacharjee2,Brandenberger2}, among others.
They could also enforce the possible
origin for the seed fluctuation density perturbations that
imprint in the cosmic microwave background radiation (CMBR), which
became the large-scale structure of the Universe  as observed today
\cite{Stebbins,Sato,Brandenberger,Valdir}.
In view of the possibility that supersymmetry was realized in the early
Universe and was broken before or at the same time as cosmic string were formed,
many recent works investigate  cosmic string in connection with 
a supersymmetric framework in inequivalent form\cite{Morris,Davis,Edelstein}.

In this work, we analyse the possibility that a D-and-F
cosmic string may be generated in a supersymmetric environment.
The mixing proposed for the gauge-field kinetic terms has already been discussed in
diferent contexts\cite{Dienes,Foot}. 
This paper is outlined as follows. In Section 2, we start by
presenting the model under consideration, as well as some of its
basic characteristics. In Section 3, we show that the model admits
a stable static cosmic string configuration. The
potential that rules the symmetry breaks is analysed  and the
ground state is read off.  We also discuss the mixing terms and
the masses of the bosonic excitations. We show that  supersymmetry
is spontaneously broken in the core. In Section 4, we discuss the
mixing terms in connection with
 the masses of the fermionic excitations;  with the help of  the
supersymmetry transformations of the component fields, we compute
the fermionic zero-modes and tackle the problem of the fermionic
charged carriers.  Next, in Section 5, we propose possible
cosmological implications of the model here analysed. Finally, in
Section 6, our Concluding Remarks are summarised.

\section{Supersymmetric extension of the $U(1)\times U(1)'$-model
and the spontaneous breaking. }

\noindent

In this section, we set up the superspace and the component-field
version of the $U(1)\times U(1)'$-gauge theory in which framework we shall
seek for a cosmic string solution. (For the algebraic manipulations with the
Grassmann-valued spinorial
coordinates and fields, we refer to the conventions adopted in the
work of Ref.\cite{Piguet}.) As  our working model, we propose the supersymmetric extension of a
$U(1)\times U(1)'$-Higgs model described by the
superspace Lagrangian:

\begin{equation}
\begin{array}{rl}
{\cal L}= &\bar \Phi_i  e^{2 q Q_i {\cal V}} \Phi_i +
\alpha_1\left(  X^\alpha  X_\alpha +
\bar X^{\dot{\alpha}} \bar X_{\dot{\alpha}}\right) \\
&+\alpha_2 \left( Y^\alpha Y_\alpha +
\bar Y^{\dot{\alpha}} \bar Y_{\dot{\alpha}}\right)+
2\alpha_3 \left( X^\alpha Y_\alpha +
\bar X^{\dot{\alpha}} \bar Y_{\dot{\alpha}}\right)\\
&+ W(\Phi) + \bar W(\Phi) + k D + \tilde k \tilde D,\label{l1}
\end{array}
\end{equation}

\noindent
where the mixing parameters, $\alpha_1 $, $\alpha_2$ and $\alpha_3$, will be connected
with  the cosmic string,  $i=1,...,N$ and the superfields
appearing herein being defined in the sequel. $D$ and $\tilde D$
are component fields accommodated in the
 superfields ${\cal A}$ and
${\cal V}$, respectively.

This model consists of a family of chiral superfields,
$\Phi_i$, that read as below:

\begin{equation}
\Phi_i(x, \theta) = \phi_i(y) +
\sqrt{2} \theta^{\alpha} \psi_{_{i}\alpha}(y) + \theta^2 F_i(y),
\end{equation}

\noindent
where $y^{\mu} = x^{\mu} + i \theta \sigma^{\mu} \bar \theta $.
The so-called vector superfield,
 ${\cal V}$, in the Wess-Zumino gauge is given by the $\theta$-expansion as follows:

\begin{equation}
{\cal V} = -\theta \sigma_{\mu} \bar \theta A^{\mu}_1(x) - i \theta^2 \bar \theta \bar \lambda_1(x) +
i \bar \theta^2 \theta \lambda_1 + \frac{1}{2} \theta^2 \bar \theta^2 \tilde D.
\end{equation}

\begin{equation}
{\cal A} = -(\theta \sigma_{\mu} \bar \theta A^{\mu}_2(x) + i \theta^2 \bar \theta \bar \lambda_2(x) -
i \bar \theta^2 \theta \lambda_2 + \frac{1}{2} \theta^2 \bar \theta^2  D),
\end{equation}

\noindent
where the gauge-field strength superfields are $
X_{\alpha} =  - \frac{1}{4} \bar D^2 D_{\alpha}{\cal V}\label{X}$
and $Y_{\alpha} =  - \frac{1}{4} \bar D^2 D_{\alpha}{\cal A}$, with
$D_{\alpha}$ and $\bar D_{\dot \alpha}$ standing for the supersymmetry
covariant derivatives.

In  component-field form, our complete Lagrangian density is split
into a bosonic piece (${\cal L}_B$),
a fermionic contribution (${\cal L}_F$), the
Yukawa part (${\cal L}_Y$) and the potential ($U$):

\begin{equation}
{\cal L} = {\cal L}_B + {\cal L}_F + {\cal L}_Y -U,
\end{equation}

\noindent where ${\cal L}_B$ and ${\cal L}_F$ are respectively the
bosonic and fermionic Lagrangian densities, ${\cal L}_Y$
encompasses all Yukawa-type couplings and U stands for the
potential:

\begin{equation}
U = \frac{\alpha_1 }{2}\tilde D^2 + \frac{\alpha_2}{2}D^2 +
\alpha_3 D \tilde D + \sum_i \bar F_i F_i. \label{pot}
\end{equation}

The Fayet-Illiopoulos  D-term provides a possible way of
spontaneously breaking SUSY\cite{Wess}. In this case, the
superpotential is

\begin{equation}
W = m\Phi_+\Phi_- .
\end{equation}

\noindent
where we replace the label $i = 1,2 $ by $i = +, -$,where $+$ and $-$ refer to the $U(1)$-charges of the superfields. These charges rule the gauge transformations of $\Phi_+$ and $\Phi_-$ . At this point, we should mention that, by virtue of the mixing between
the auxiliary fields $D$ and $\tilde D$ along with the Fayet-Illiopoulos (F-I)
terms in (\ref{l1}), we are able to attain a superconducting cosmic string configuration
with  4-chiral superfields, without the need of introducing of  an extra neutral
matter supermultiplet:

\begin{equation}
\begin{array}{ll}
D = \frac{1}{\beta}\left[\frac{\alpha_3}{2}\sum_i qQ_i |\phi_i|^2 +
\alpha_3\tilde k -\alpha_1 k \right],\\
\\
\tilde D =\frac{1}{\beta}\left[- \frac{\alpha_2}{2}\sum_i q Q_i |\phi_i|^2   + \alpha_3 k - \alpha_2 \tilde k \right],\\
\\
\bar F_i  = - \frac{\partial W}{\partial \bar \phi_i}= - m  \phi_i,
\end{array}
\end{equation}

\noindent
with $\beta = \alpha_1 \alpha_2 - \alpha_3^2$.
Now, let us analyse the positivity of the potential. For this, 
let us  write the D-and-F terms in the Lagrangian as below:

\begin{equation}
-U = \frac{1}{2} {\cal D}^t M {\cal D} + \frac{1}{2}{\cal D}^t K + \frac{1}{2}K^t {\cal D} - \sum_i \bar F_i F_i \label{posit}
\end{equation}

\noindent
where ${\cal D} = \left(\begin{array}{ll}D \\
\tilde D
\end{array}\right)$, $K = \left( \begin{array}{ll}k\\
\tilde k \end{array}
\right) $ and $ M = \left(\begin{array}{ll}
\alpha_1 & \alpha_3\\
\alpha_3 & \alpha_2 \end{array}\right), $

\noindent
where the latter is positive-definite.
By adopting of the redefinition, $\tilde {\cal D} = {\cal D} + M^{-1} K $ , we find that

\begin{equation}
U = \frac{1}{2} K^t M^{-1} K + \sum_i \bar F_i F_i,
\end{equation}

If we consider the F-term, and we choose $\tilde k\neq 0$ and $ k \neq 0$, 
the potential (\ref{pot}) can be split as below:

\begin{equation}
\begin{array}{lll}
U_{cs}& =& \frac{1}{\beta^2 }\left[ \frac{\alpha_2}{8} q^2 (|\phi_+|^2 - |\phi_-|^2)^2\right.\\
&&+ \left(
\beta m^2 + \frac{\alpha_2}{2}q\tilde k  - \frac{\alpha_3}{2}q k  \right)|\phi_+|^2  \\
&&+\left(\beta m^2
-\frac{\alpha_2}{2}q \tilde k + \frac{\alpha_3}{2}q k\right)|\phi_-|^2  \\
&&\left.+\frac{\alpha_1}{2} k^2 +\frac{\alpha_2}{2}\tilde k^2 - \alpha_3\tilde k k  \right],
\end{array}
\label{potcs}
\end{equation}

Now, we analyze the delicate issue of gauge
symmetry and supersymmetry breakings, and the consequent formation
of a cosmic string configuration. By minimizing
the potential of eq.(\ref{pot}), we shall focus on
the possibility of finding a supersymmetric 
cosmic string.

For our purposes, we can work with the potential to cosmic string  
as given below:

\begin{equation}
\begin{array}{lll}
 U(\phi_+, \phi_-) & = & \frac{\lambda_{\phi_+}}{4} (
|\phi_+ |^2 - \eta^2)^2 - f |\phi_+ |^2|\phi_- |^2 +
\\ && +
\frac{\lambda_{\phi_-}}{4}|\phi_- |^4 +
\frac{m^2_{\phi}}{2}|\phi_-|^2,
\end{array}
\end{equation}

\noindent
with $\lambda_{\varphi^+} = \frac{q^2}{2\beta }$, $\lambda_{\varphi^-} = \frac{q^2}{2\beta }$,
$f = \frac{q^2}{4\beta}$ and $m_{\phi}^2 = \frac{4\alpha q k + 4q^2 \eta^2}{4\beta}$.

A minimum vacuum configuration for a static vortex may come out if
we have $\phi_+=\eta$, $\phi_- = 0$ and  $ \eta^2 =  \frac{2 k}{q}v
$  where $v = \sqrt{[1 - \frac{2\tilde k}{k}(\alpha k - \frac{\tilde k}{2})]} $  and  in potential
(\ref{potcs}) the $m^2$-parameter is given by

\begin{equation} 
m^2= \frac{q \alpha k}{2\beta}(1 - \frac{v + \tilde k/k}{\alpha}).
\end{equation}

\noindent
with $\alpha_1 = \alpha_2 =1$ and $\alpha_3 = \alpha$ with 
$\beta = 1-\alpha^2 $.

Another  important feature of the cosmic
string configuration regards its core. It is described  by
$\phi_+=\phi_- =0$.
The U(1)-gauge symmetry
is exact and the U(1)'-gauge symmetry is broken.
Supersymmetry is broken in the  core.

We note that the
cosmic string only exist to $m^2>0$, then we have that $k>0$ with 
$ \tilde k < (\alpha - v) k$, that can be fine-tune ajust. But 
the important analyse is tha $k \neq 0$ and $\tilde k \neq 0$ to
we have the cosmic string potential with D-and-F-term.

Neverthless, the final comment is that our choice for $m$ does eliminate the  
flat directions from our model, that
with appear when $m  =0$ [to review about the flat directions
see\cite{Zumino}]. So, we are sure about  stability of our
string configuration.

With our potential, despite the fact that we have two $U(1)$'s at
play, the flat directions disappear whenever the parameter $m$
is non- vanishing. Usually, the flat directions
inherent to supersymmetric section gauge theories appear because a
single $U(1)$-symmetry factor is involved \cite{Zumino}. Our
model, however, is based on 2 Abelian factors and this is crucial
to ensure that, if only $m$  were non-vanishing, the
flat directions problem would be bypassed. Nevertheless, as a final
comment we mention that, even in presence of eventual flat directions,
stability would not be jeopardized, since the 1-loop corrections do not
change the conditions for the SUSY breaking.

\section{Bosonic cosmic string configuration }

\noindent

In this section, we analyse the possibility of obtaining a
stable bosonic configuration  for a  cosmic string
in our supersymmetric approach. We  notice that a mixing term can
only occur when there are two or more field-strength tensors,
$F^1_{\mu \nu}$, $F^2_{\mu \nu}$. This only arises for Abelian
groups. Thus,  the simplest gauge model  with  a   mixing in the
kinetic terms is a   model with gauge group $U(1)\times
U(1)'$. Let us denote the field strengths of the two $U(1)$-fields
by $F_{\mu \nu}$ and $H_{\mu \nu}$, respectively.

In the supersymmetric version\cite{Dienes}, this mixing appears naturally given
by

\begin{equation}
-\frac{1}{4} F^1_{\mu\nu}F_1^{\mu\nu} -
\frac{1}{4} F^2_{\mu\nu}F_2^{\mu\nu}
-\frac{\alpha}{2} F^1_{\mu\nu}F_2^{\mu\nu}.
\end{equation}

The constant $\alpha$ is a physical parameter and cannot be
completely scaled away in the presence of interactions, as it shall
discussed in the next section.
Now, we discuss possible string configurations in connection with the
choice of basis in field space. We get two possibilities
for cosmic strings.

In this section, we consider the case where the
kinetic term can be diagonalised by performing the orthogonal
transformation,

\begin{equation}
\begin{array}{ll}
H^{\mu} = A_1^{\mu} + \alpha A_2^{\mu}\\
A^{\mu} = \sqrt{1- \alpha^2} A_2^{\mu}.
\end{array}\label{trans2}
\end{equation}

The requirement of a positive kinetic energy implies that $|\alpha |
< 1$. Though the diagonalisation of eq. (\ref{trans2}) eliminates
the mixing term, the effect of this mixing is present in the couplings, after
the field redefinitions $A_{\mu}$ and $H_{\mu}$ are adopted. This
means that the elimination of the mixing has a physical implication: it
yields a relative strength between the coupling constants that govern
the interactions of $A^{\mu}$ and $H^{\mu}$ with matter. This is why, as we have
anticipated above, the $ \alpha $-parameter cannot be completely
absorbed into field reshufflings.
We replace these  transformations in the bosonic Lagrangian,

\begin{equation}
{\cal L}_B = D^i_{\mu}\phi_i \bar D_i^{\mu} \bar \phi^i 
-\frac{1}{4} F_{\mu\nu}F^{\mu\nu} - \frac{1}{4}H_{\mu\nu}
H^{\mu\nu} - U,
\label{B1}
\end{equation}

\noindent
where the gauge-covariant  derivatives read as follows:

The gauge-covariant derivatives take the form:

\begin{equation}
D_{\mu}^i = \partial_{\mu} + i \frac{q}{2} Q^i H_{\mu} -
i \frac{\alpha q}{2 \sqrt{\beta} }Q^i A_{\mu}. \label{covariant2}
\end{equation}

The field-strengths are defined as usually:
$F_{\mu \nu} = \partial_{\mu}A_{\nu} - \partial_{\nu} A_{\mu}$ and
$H_{\mu \nu} = \partial_{\mu}H_{\nu} - \partial_{\nu} H_{\mu}$, with $A_{\mu}$
and $H_{\mu}$ being the gauge fields.

Our ansatz for the supersymmetric generalization of the
Nielsen-Olesen string configuration \cite{Nielsen} is proposed as follows:

\begin{equation} \begin{array}{ll}
\phi_+ = \varphi_+(r )e^{i\theta}\\
\phi_- = \varphi_- (r) e^{-i\theta}\\
H_{\mu} = \frac{2}{q }(H(r)-1)\delta^{\theta}_{\mu},
\end{array}\label{vortex1} \end{equation}

\noindent parametrised in cylindrical coordinates
$(t,r,\theta,z)$, where $r\geq 0$ and $0 \leq \theta < 2 \pi $,
with $H_{\mu}$ as the gauge field. The boundary conditions for the
fields $\phi_{\pm}(r) $ and $H(r)$ are the same as those for the
ordinary cosmic strings:

$$ \begin{array}{lll} \varphi_+(r) = \eta , &\varphi_-(r) =0& r
\rightarrow \infty, \\ \varphi_+(r) =0, &\varphi_- (r)=0& r
\rightarrow 0,
\end{array}
$$

\begin{equation}
\begin{array}{ll} H(r) = 0 & r \rightarrow \infty \\ H(r) =1 & r
\rightarrow 0;  \end{array} \label{config1}
\end{equation}

\begin{equation} 
A_{\mu} = 2\sqrt{\beta}A(r)\delta_{\mu}^{z}.
\label{vortex2}
\end{equation}

\begin{equation}
\begin{array}{ll} A(r) =
0 & r \rightarrow \infty \\ A(r) = c & r = 0.
\end{array}\label{config2}
\end{equation}

The Euler-Lagrange equations for the $\phi_{\pm}$-fields are given by

$$
\begin{array}{ll}
\varphi_{+}'' + \frac{1}{r}\varphi_{+}' &- \varphi_{+}[H^2 +
\frac{\alpha^2 q^2}{r^2}A^2 + \lambda_{\phi_+} ( \varphi_+^2 -
\varphi_-^2 ) =0,
\end{array}
$$

\begin{equation}
\begin{array}{ll}
\varphi_{-}'' + \frac{1}{r}\varphi_{-}' &+ \varphi_{-}[H^2 +
\frac{\alpha^2 q^2}{r^2}A^2 - \lambda_{\phi} ( \varphi_+^2 -
\varphi_-^2 ) =0.
\end{array}
\end{equation}

Now, the equations for the gauge fields read as:

\begin{equation}
H'' - \frac{1}{r}H' + q^2 (\varphi^2_+ - \varphi^2_-)H =0,
\label{equa1}
\end{equation}

\begin{equation}
A'' + \frac{1}{r}A' + \frac{q^2\alpha^2}{4\beta } (\varphi_+^2- \varphi_-^2)
A = 0.
\end{equation}

We choose the basis given by (\ref{trans2}). These particular combinations render
chearer the discussion of the fermionic current and the breaking of SUSY in the string core.
The extra gauge field plays a crucial role in connection
with the breaking of the gauge symmetry outside the string core.

\section{Fermionic current-carrying cosmic string}

At this stage, some highlights on the fermionic configurations are worthwhile.
This issue has already been discussed in a previous paper
\cite{Cris1}. All we have done in the previous sections concerns the
bosonic sector of the $U(1)\times U(1)'$-theory; to introduce
the fermionic modes, which have partnership
 with the vortex configurations,
we take advantage from SUSY invariance. By this, we mean that the
non-trivial configurations for the fermionic degrees of freedom
may be found out by acting with SUSY transformations on the
bosonic sector. In a paper by Davis et al. \cite{Davis}, this
procedure is clearly stated  and we follow its details here.

The SUSY transformations of the component fields for the
$U(1)\times U(1)'$-model may be found in  ref. \cite{Wess}, where
the component-field transformations of the matter and gauge
supermultiplets are explicitly written down. Going along the steps
presented in the work of ref.\cite{Cris1}, where SUSY
transformations act upon the string bosonic configuration, we get
the fermionic zero-mode configurations which turn out to be:

\begin{equation}
\begin{array}{ll}
\psi^{\pm}_a = i \sqrt{2}\sigma^\mu_{a
\dot a}\bar \varepsilon ^{\dot a}D_{\mu} \phi^{\pm} + \sqrt{2} \varepsilon_a F^{\pm},\\
\xi _a = \sigma^{\mu \nu}_{a\dot a}\varepsilon^{\dot a}H_{\mu \nu}
+ i \varepsilon_a \tilde D,
\end{array}\label{ferm2}
\end{equation}

\begin{equation}
 \lambda _a = \sigma^{\mu \nu}_{a\dot a}\varepsilon^{\dot a}F_{\mu \nu} +
i \varepsilon_a  D.\label{ferm4}
\end{equation}

It is worthwhile to notice that the SUSY transformations lead to a
vector supermultiplet ($\cal V $) that is no longer in the
Wess-Zumino gauge; to reset such a gauge for $\cal V$, we have to
supplement the SUSY transformation by a suitable gauge
transformation that has to act upon the matter fields as well. All
these facts have already been taken into account in the zero-modes
of eqs.(\ref{ferm2})-(\ref{ferm4}).

Now, we come to the question of the mass for the  fermionic excitations.
The gaugino-mixing fermionic term reads as:

\begin{equation}
\begin{array}{ll}
 -i {\lambda_1^{\alpha }}
\sigma^{\mu}_{\alpha \dot\alpha } \partial_{\mu} \bar {
\lambda_1^{\dot \alpha }} -i  \lambda_2^{\alpha}
\sigma^{\mu}_{\alpha \dot \alpha }\partial_{\mu} \bar
\lambda_2^{\dot \alpha }\\
- {\lambda_1^{\alpha}} \sigma^{\mu}_{\alpha \dot
\alpha}\partial_{\mu} \bar \lambda_2^{\dot \alpha} -i \alpha
\left( \lambda_2^{\alpha } \sigma^{\mu}_{\alpha \dot \alpha
}\partial_{\mu} \bar {\lambda_1^{\dot \alpha }}  \right).
\end{array}
\end{equation}

The field reshufflings below diagonalise the fermion mass matrix:

\begin{equation}
\begin{array}{ll}
\xi^\alpha = \lambda_1^\alpha  + \alpha \lambda_2^\alpha ,\\
\lambda^\alpha  = \sqrt{\beta } \lambda_2^\alpha .
\end{array}\label{trans1}
\end{equation}

With this field basis, we  obtain the fermionic Lagrangian:

\begin{equation}
{\cal L}_F = i \psi_i \sigma^{\mu}D^i_{\mu}\bar \psi_i 
-i  \xi^{\alpha}
\sigma^{\mu}_{\alpha \dot \alpha }\partial_{\mu} {\bar {
\xi^{\dot \alpha }}} -
i {\lambda^{\alpha }}
\sigma^{\mu}_{\alpha \dot\alpha } \partial_{\mu} \bar
{ \lambda^{\dot \alpha }},
\label{F1}
\end{equation}

\noindent
with the covariant derivative reading as

\begin{equation}
D_{\mu}^i = \partial_{\mu} + i \frac{q}{2} Q^i H_{\mu} -
i \frac{\alpha q}{2 \sqrt{\beta} }Q^i A_{\mu},
\end{equation}

\noindent
and the Yukawa potential

\begin{equation}
{\cal L}_Y =  - \frac{i q}{\sqrt{2 }}Q_i\left[\phi_i \bar \psi_i ({\bar {\xi}}-
\frac{\alpha}{\sqrt{\beta}}{\bar{ \lambda}})   -
\phi_i \psi_i  (\xi - \frac{\alpha}{\sqrt{\beta}}\lambda ) \right].
 \label{Y1}
\end{equation}

We propose our fermion solutions as given by:

\begin{equation}
\psi_\pm  =\psi_i(r,\theta) e^{\zeta(z,t)};
 \hspace{.3 true cm}  \xi =\xi(r,\theta) e^{-\zeta(z,t)},
\end{equation}

Notice that the left-moving superconducting current presents the function
$\zeta(z,t)$, which is taken  the
same for  all fermions.

The symmetry breaking also modifies the spinor mass terms; in this case, we
 have, after the breaking, the following fermionic mass term:

\begin{equation}
{\cal L}_{mass} =   \frac{iq \eta}{\sqrt{2 }}\left[\bar \psi_+ ({\bar {\xi}}-
\frac{\alpha}{\sqrt{\beta}}{\bar{ \lambda}})   -
\psi_+  (\xi - \frac{\alpha}{\sqrt{\beta}}\lambda ) \right] \label{fmlagrangian}\end{equation}

In fact, we see that only $\psi_+$ acquires mass, because $\phi_-=0$ in the vacuum.
In this form, this model describes one Goldstone fermion, one scalar and a vector field with mass
$\frac{\alpha}{\sqrt{2\beta }}q \eta$.  Notice that these masses fit the
well-known supersymmetry mass formula for the spontaneously broken
case \cite{Girandelo}.

Before  the cosmological application,
we study the currents of fermionic particles. 
For this we shall consider our string to be describable 
by means of a surface action and accordingly integrate 
the action  over the transverse  degrees of freedom, i.e.
$(r, \theta)$-coordinates (to review of the procedures see \cite{Vilenkin1}).
In this case let us conside by use of the
simmetry of the  problem that $A_{\mu}(\vec{x}) = A_a(t)$, where $a = z, t$.
That massless
fermions interact with the electromagnetic gauge field in the core. The 
gauge field, $A_z$, as we saw, is z-independent; it couples to the fermionic particles of
 the matter supermultiplet. The latter are massless in the core and massive
outside, as given by eq. (\ref{fmlagrangian}). The analysis of the fermionic
current is better carried out in the 2-dimensional sheet. 
There is a  current induced in a z-directed string by a homogeneous
electric field, ${\bf E}$. The symmetry of the problem suggests that the current $J^a$ can be
also z-independent. Then,

\begin{equation}
\frac{dJ_z}{dt} = \frac{\alpha^2}{2\beta } q^2 E_z
\end{equation}

So, we can interpret that the electrical field $E_z$ is the
responsible for the orientation of the charges in the z-direction.

\section{Possible Cosmological Implications of a Supersymmetric Superconducting Cosmic String}

\noindent

In this section, we point out that cosmic strings  associated with phase
transitions in our model can, through their collapse, annihilation,
or other processes, be sources for
heavy fermions  with masses in the range $10^{11}$ to $10^{13}$ GeV, whose
decay products may be observable in the Universe today.

Recent works
have investigated  extragalactic $\gamma$-rays and ultra-high
energy cosmic rays
in connection with supersymmetric cosmic strings\cite{Bhattacharjee,Kachelrie,Berezinsky}.
In this approach, the latter may
be sources of  Higgs particles with mass comparable
to the (explicit) supersymmetry breaking scale (TeV), and
superheavy gauge bosons of
mass of order $\eta$, where $\eta $ might be of order of the GUT scale.

In view of the scenario mentioned above, we propose here that
our supersymmetric superconducting cosmic string, for which
supersymmetry
is spontaneously broken in the core,
may be source of  fermionic particles with mass comparable
to the spontaneous supersymmetry breaking range
$\sim 10^{11}$ to $10^{13}$ GeV.
The gauge symmetry breaking is  usually associated
to massive particle production\cite{Vilenkin1}; in our approach, where the string becomes
superconducting, the fermionic carriers responsible for superconductivity
may play a significant role in the process of massive particle production.

In this part of the work, let us consider the fermions that carry charges under both
the original $U(1)$ which is spontaneously broken and the electromagnetic
$U(1)'$ which remains unbroken. For a particle  of
charge $q$,
indeed, consider the current induced in the z-directed string by a
homogeneous (z-independent)
electric field, $E_z$.
The string develops an electric
current which grows in time,

\begin{equation}
\frac{d J_z}{dt} \sim q^2 \frac{\alpha^2}{2\beta} E_z, \label{current}
\end{equation}

\noindent
where $E_z$ is the component along the string.
We analyse the case of fermionic supercondutivity
 in our model; the charged fermions, $\psi_i $, and the neutral fermions,
$\lambda $, acquire their mass as a result of gauge
symmetry breaking, which is responsible for the string formation; then, they
are massive outside, but massless inside the string, as  in
(\ref{fmlagrangian}). These fermions are electrically charged,
then the strings have
massless charged carriers which travel along the string at the speed of light,
as given by (\ref{current}). This current in the core is generated by the
supersymmetry breaking  scale,
and Goldstone fermions appear. The fermion mass outside the string is

\begin{equation}
M_F = g \eta, \label{fmass}
\end{equation}

\noindent
where
$g = q \alpha $ is the Yukawa coupling of the fermion
to the Higgs field of the string,  given by (\ref{fmlagrangian}), and $\alpha $ is the
fine structure constant that fits the experimental data.

If we consider that
loops may be formed by interaction between our long cosmic strings,
we have that, with fermionic
charge carriers, they are expected to eject, in their late stages, high-mass
particles which  subsequently decay to produce ultra-high cosmic rays,
neutrino, and hadronic radiation (particles inside the string can be 
thought of as a one-dimensional Fermi gas) \cite{Hill}. When an electric
field is applied, the Fermi momentum grows as
$\frac{d p_F}{dt} = \pi q E $,
where $p_F$ and the number of  fermions per unit length\cite{Hill},
$ N^F = L \frac{p_F}{\pi }$, also grows:

\begin{equation}
\frac{dN^F}{dt} \sim q  E L.\label{fnumber}
\end{equation}

Now, that the constant are compatible, we can use
(\ref{current}) to get

\begin{equation}
J^F= q\frac{\alpha}{\sqrt{2\beta }L } N^F.
\end{equation}

\noindent
The current grows until it
reaches a critical value,

\begin{equation}
J^F_C \sim q \frac{\alpha}{\sqrt{2 \beta}} M_F,
\end{equation}

\noindent
where

\begin{equation}
p_F = M_F = q \frac{\alpha }{2\beta} \eta . \label{beta1}
\end{equation}

At this point, particles at the Fermi level have sufficient
energy to leave the string. The fermion mass, $M_F$, does not exceed the supersymmetry  breaking
scale, which is compatible with
gauge symmetry vortex formation  around  $10^{16}$GeV. This result is compatible with the
supersymmetry breaking in the core and with the fermion solutions given by the supersymmetry
transformations (\ref{ferm2}),i.e., the solutions to $\psi$ and $\xi $ fall off to
zero at infinity. Hence,

\begin{equation}
J_C < J_{max} \sim q \frac{\alpha}{2\beta}  M_F.\label{jmax}
\end{equation}

The value of  $\alpha $ may be expected to be less than 1,
with $q^2 \sim 10^{-2}$, because the $\beta \sim 1$ \cite{Vilenkin1}
and  $M_F$ and $\eta $ may take values as below:

$$
\begin{array}{lll}
M_F \sim 10^{13} GeV, & g \sim 10^{-3}, &\alpha \sim  10^{-2} \\
M_F \sim 10^{12} GeV,& g \sim  10^{-4}, & \alpha \sim 10^{-3}\\
M_F \sim 10^{11} GeV,& g \sim  10^{-5}, & \alpha \sim 10^{-4}.
\end{array}
$$

It is a major issue to justify, on the basis of first principles,
the way our parameters can be chosen with the values displayed
above. First of all, the fact that $M_F$ may be taken in the range
$10^{11} $ to $10^{13} $ GeV is supported by our knowledge that,
if SUSY is broken and the gravitino mass is to be of the order of
TeV, then the SUSY breaking scale lies at the intermediate scale
between $10^{11}$ to $10^{13}$ GeV. As for the $\beta $-parameter,
the values we have chosen are compatible to reproduce cosmic
string phenomenology. On the other hand, being SUSY spontaneously
(and, therefore, softly) broken, stability against quantum
correction legitimates the values proposed in our calculations
above. Then, we can state that it is supersymmetry and its power
in the renormalization programme the main support in favor of our
proposal of parameter choices.

Consequently, in this simplified picture, the growth of the
current ends at $J_C$ and the string starts producing particles at
the rate (\ref{fnumber}). In our  model, we may accommodate the
production of extremely energetic photons, quarks and neutrinos
through decays mediated by  superheavy bosons (typically of
GUT-scale mass $10 ^{16} $ GeV). We may also include the
production of heavy fermions with mass  in the range  $10^{11}$ to
$10^{13}$ GeV, compatible with spontaneous supersymmetry breaking
as discussed in this section.

\section{Concluding Remarks}

\noindent

The goal of this work is the investigation of the structure of a  cosmic
string in the framework of a supersymmetric Abelian Higgs model. We show that it is
possible to set a $U(1)\times U(1)'$-Lagrangian with the property of a fermionic
current in the core of the string generated by a D-term and the superpotential $W$. 
Supersymmetry is spontaneously
broken in the core, while it is preserved outside the string.
We found that the introduction of a mixed kinetic term involving the
two Abelian factors appears as the only way to custruct the bosonic cosmic stiring
with D-and F-term. 

The propagation of
these fermionic excitations should be understood in connection
with SUSY breaking and gives us an  interesting result on the
zero-modes related to the field that is not confined to the
string. There are interesting questions that remain to be
contemplated, concerning supersymmetric topological defects and
their implication to Particle Physics in a supersymmetric version
of the extended Standard Model, with a very general Lagrangian
based on $SU(2) \times U(1) \times U(1)'$ \cite{Babu}, whose
phenomenological implications have been currently discussed in the
literature.

Finally, our discussion on a possible cosmological scenario for
supersymmetric cosmic strings with superconductivity opens up a
viable way of trying to fit the phenomenology concerning the
production of extragalactic $\gamma $- rays and ultra-energetic
cosmic rays. In fact, the fermionic current inside the vortex
which can give us  supermassive particles outside the string, is
given by a  supersymmetry beaking as justified in the last
section; the connection between the supersymmetry spontaneous
breaking scales ($\sim 10^{11}$ to $10^{13}$ GeV) with massive
particles in our model is given by the $\beta $- mixing parameter.
These supersymmetric fermionic particles may have our product
detectable in Fly's Eye and its successor HiRes, as well as in the
Akeno Giant Air Shower Array  (AGASA)
experiments\cite{Kachelrie1}. Interesting investigation about our
model may be done in future works \cite{Portella} using the
fragmentation method for these supermassive particles. The
air-showers  produced by primaries can be studied from
fragmentation of the X-supersymmetric particles as it may be found
in Ref:\cite{Kachelrie}. An interesting issue is to perform the
Feymman expansional operator  method\cite{Pi} that can be extended
to include  other contributions, such as earthy curvature etc..
So, our approach is a mechanism that can give us the possibility
to obtain these energies if the scale for cosmic string scale
formation is of the order of $ 10^{16}$ GeV.  This issue may offer
a very rich set up to probe the details of our model.

\vspace{.7 true cm}

{\bf Acknowledgments:}

The authors would like to thank (CNPq-Brasil)  for the invaluable financial
support. C.E.C Lima is acknowledged for a careful reading of the manuscript.

\end{document}